# A Depth-Adaptive Filtering Method for Effective GPR Tree Roots Detection in Tropical Area

Wenhao Luo, *Graduate Student Member, IEEE*, Yee Hui Lee, *Senior Member*, *IEEE*, Mohamed Lokman Mohd Yusof and Abdulkadir C. Yucel, *Senior Member, IEEE*

*Abstract*— This study presents a technique for processing Step-frequency continuous wave (SFCW) ground penetrating radar (GPR) data to detect tree roots. SFCW GPR is portable and enables precise control of energy levels, balancing depth and resolution trade-offs. However, the high-frequency components of the transmission band suffers from poor penetrating capability and generates noise that interferes with root detection. The proposed time-frequency filtering technique uses a short-time Fourier transform (STFT) to track changes in frequency spectrum density over time. To obtain the filter window, a weighted linear regression (WLR) method is used. By adopting a conversion method that is a variant of the chirp Z-Transform (CZT), the time-frequency window filters out frequency samples that are not of interest when doing the frequency-to-time domain data conversion. The proposed depth-adaptive filter window can self-adjust to different scenarios, making it independent of soil information and effectively determines subsurface tree roots. The technique is successfully validated using SFCW GPR data from actual sites in a tropical area with different soil moisture levels, and the two-dimensional (2D) radar map of subsurface root systems is highly improved compared to existing methods.

## I. INTRODUCTION

TREE root system analysis is a reliable method of understanding the trees' state of health. One essential way to monitor trees' condition is to map the structure of the tree root system, knowing the position of the tree roots, estimating their structure, direction of extension, dimension and material properties [1-3]. In recent years, ground-penetrating radar (GPR) has been employed in the field of root detection because of its advantage of high-resolution, high efficiency, and continuity [4, 5].

One classical GPR is the time domain radar that emits extremely short pulses, typically within nanoseconds, or even less for a large bandwidth signal. Pulse antennas with a fixed center frequency are usually used to adapt to different depth and resolution requirements, as a trade-off between each other: The higher frequency signals provide higher resolution with limited penetration capability, while the lower frequency signals provide lower resolution with high penetration capability. When the target's location and dimension are unknown, subsurface target detection becomes a challenge. In order to accurately map subsurface tree roots, multiple methods have been carefully studied and proposed.

Some researchers introduced the multi-frequency GPR in order to provide both good penetration depth and high-resolution images of subsurface targets. A multi-frequency GPR survey to detect ice and sediment layers in Svalbard permafrost is detailed in [6]. However, no data fusion method was used. In [7], the authors proposed a simple averaged sum of the inverted tomograms of the low- and high-frequency GPR survey, which overlaps with multi-frequency GPR surveys in the spectral domain. Liu concluded that extrapolation by deterministic deconvolution (EDD) is effective in taking advantage of multi-frequency data [8, 9]. Zhao and Lu [10] proposed a multi-frequency GPR data fusion algorithm based on a time-varying weighting strategy. Though the multi-frequency GPR data fusion methods introduced above are effective in improving the GPR imaging and interpretation, these methods are not efficient, and/or difficult-to-implement, and/or not suitable for tree root detection. The need to use multiple central frequency antennas corresponding to different frequency ranges for data collection makes these methods impractical for implementation. This makes instrumentation inconvenient and the repeated experiments to ensure the same reference point make experimentation laborious. All these multi-frequency fusion methods need to be improved in order for ease of implementation and automatic data fusion.

Step-frequency continuous wave (SFCW) GPR with ultra-wide-band (UWB) frequency samples and synchronous energy control outperforms the traditional central frequency pulse GPR [11, 12]. The SFCW radar offers high-resolution imaging and deep subsurface penetration [13], but the use of high frequencies is limited to shallow layers since the high frequencies tend to be below the receiver's noise level at deeper depths, especially in tropical regions with high soil moisture content. Several studies have been proposed to deal with the above shortages of SFCW GPR. Federico and Maurizio [14] changed the size of the Inverse Fast Fourier Transform (IFFT) operator to characterize agricultural soil morphology. They divided the total bandwidth of SFCW data into multiple

Manuscript received February 17, 2023; revised April 24, 2023; accepted May 24, 2023. This work was supported by the Ministry of National Development Research Fund, National Parks Board, Singapore. (Corresponding authors: *Yee Hui Lee; Abdulkadir C. Yucel*.)

Wenhao Luo, Yee Hui Lee, and Abdulkadir C. Yucel are with the School of Electrical and Electronic Engineering, Nanyang Technological University, Singapore 639798 (e-mail: wenhao.luo@ntu.edu.sg; eyhlee@ntu.edu.sg; acyucel@ntu.edu.sg).

Mohamed Lokman Mohd Yusof is with National Parks Board, Singapore 259569 (e-mail: mohamed_lokman_mohd_yusof@nparks.gov.sg).

narrower frequency bands and found the effective contribution of certain features generated by certain frequency bands. However, the details of the manual choice of frequency bands and its relationship to the improvement of the images were not discussed and need further investigation. Jacopo et al. [15] generated several time-domain B-scan images with different frequency bands, and combined the visually optimal time slices from each B-scan together to get the final improved image. However, this method has the problem of continuity in the final time-domain image. Jacopo and Helge then proposed a method called Inverse Selective Discrete Fourier Transform (ISDFT) that provides a depth-dependent frequency window where the bandwidth is continuously varied to match the targets' reflection energy contents [16]. However, this method requires the characteristic attenuation values of the medium to be known in advance in order for it to be effective, which makes it a soil-information-dependent method.

To overcome the challenges of the SFCW GPR system, to enhance the time-domain radar image, we propose a depth-adaptive time-frequency filtering (DATFF) algorithm based on joint time-frequency domain analysis of GPR data and weighted linear regression (WLR) method. Joint time–frequency analysis has been successfully applied to discriminate buried objects using a UWB SFCW GPR [17]. The joint time-frequency domain analysis method, short-time Fourier transform (STFT), tracks the changes in frequency spectrum density over time. It has been widely applied to characterize construction in the GPR field [18, 19]. Recently, the feasibility of the application of STFT in tree root detection has been preliminarily explored in [20, 21]. The WLR, a well-known fitting method [22], is adopted to fit the boundary of the density map of the time-frequency spectrum data. By adopting STFT and WLR, a time-frequency filtering window is obtained. This filtering window then works together with a variant of the chirp Z-Transform (CZT) [16, 23], to be used as a conversion tool from frequency domain data to time-domain data. The filtering algorithm is applied to GPR data acquired from mature trees in different areas in Singapore. The bandwidth of the UWB signal at different time intervals has been narrowed automatically after the filtering procedure. The results show that the filtering scheme improves the imaging of tree roots and is independent of the roots' surrounding environment.

Compared with the data fusion methods using multi-frequency GPR proposed in [5-9], our method uses SFCW GPR, the advantage of using SFCW is that it can provide good penetration depth and high resolution at the same time, without repeated experiments. Based on this advantage of SFCW GPR, our method reduces unwanted high band noise and improves B-scan image quality. Moreover, the whole GPR system is portable and applicable for real-time tree root detection. Compared with the manual selecting study in [14] using SFCW GPR, our method provides a depth-adaptive frequency band selection with the change of penetrating depth over different inhomogeneous soil types. Furthermore, the DATFF generates continuous B-scan radargrams, which is an improvement over the method in [15]. Finally, this work is an independent method for enhancing B-scan radargrams when the soil properties are unknown, outperforming the ISDFT method [16]. When the properties of the transmitting soil are unknown, the proposed DATFF method can adjust the appropriate time-frequency filter window to optimize the B-scan image, while the ISDFT results will be poor when the wrong soil attenuation parameters are used. The main concept of the baseline works and our proposed method are listed in the Table I.

The rest of the paper is organized as follows: Section II describes the experimental location and equipment, as well as some basic data pre-processing methods. The details of the theories and methods we used and proposed are introduced in Section III. The results obtained by the proposed method are presented in detail, and the soil-information-independent characteristic of DATFF is discussed in Section IV. Finally, conclusions are drawn in Section V.

TABLE I
SUMMARY OF THE BASELINE WORKS AND THE PROPOSED METHOD

| GPR Type | Ref. | Authors | Main Concept |
|---|---|---|---|
| Time-domain center frequency GPR | [6] | O. Brandt et al. | Ice and sediment layers studied with multi-frequency pulse GPR; signals from different frequencies studied separately. |
| | [7] | F. Soldovieri et al. | Averaged sum of inverted tomograms from low- and high-frequency GPR survey proposed. |
| | [8][9] | L. Liu et al. | EDD, a multi-frequency data analysis method, was proposed. |
| | [10] | W. Zhao et al. | A time-varying weighting strategy-based multi-frequency GPR data fusion algorithm was proposed. |
| SFCW GPR | [14] | F. Lombardi et al. | Total bandwidth of SFCW data divided into narrower frequency bands to determine effective contribution of features generated by specific frequency bands. |
| | [15] | J. Sala et al. | Several time-domain B-scan images with different frequency bands generated and visually optimal time slices from each combined to obtain a final improved image. |
| | [16] | J. Sala et al. | ISDFT, a depth-dependent frequency window, was proposed, requiring manual input of characteristic parameters of the propagation medium. |
| | N.A. | W. Luo et al. | DATFF, a depth-adaptive frequency-selective processing method, was proposed, self-adjust to different scenarios, without requiring knowledge of the transmission medium. |

## II. THEORY AND METHODOLOGY

This chapter introduces how the frequency-domain data $S_{21}$ obtained through the SFCW GPR system test are processed by the proposed adaptive time-frequency filtering method after preprocessing. The main purpose of data filtering is to improve and simplify the interpretation of UWB frequency domain data, thus providing better time-domain B-scan data for further analysis. By using the proposed data processing method, the complex UWB frequencies in the GPR data can be precisely selected (keeping frequency components reflected by targets) and discarded (remove frequency components from noise). An ideal signal after filtering should consist of information from the full/over-half band frequency signal at the shallow depth, and information from the low-frequency signal at the deep depth. The detail of the proposed methodology is introduced in the following subsections.

A flow chart of the proposed approach is shown in Fig. 1. The pre-processed B-scan data are sent through a series of processing in order to determine the filtering window. The time-frequency domain analysis using STFT is performed on the pre-processed B-scan. After which, histogram of counts of points in time-frequency domain is generated, and then, the filtering window is fitted using WLR based on the boundary of the histogram generated. Finally, the filtering window is applied to a variation of CZT method to convert frequency domain GPR data to time domain B-scan radar map. Details are introduced as follows.

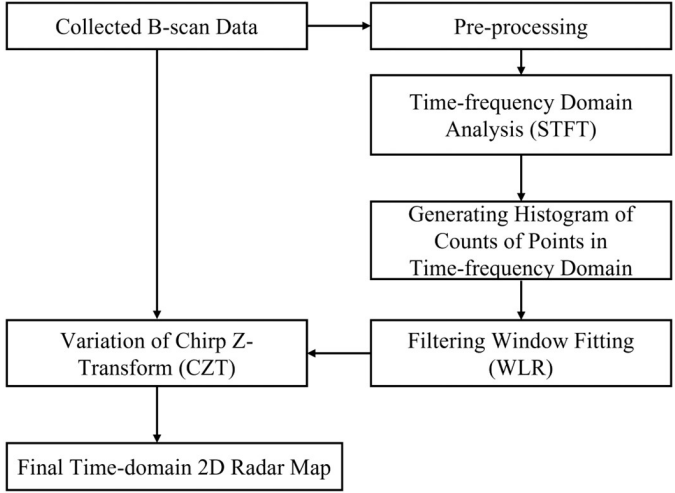

Fig. 1. flow chart of the proposed approach.

*A. Frequency to Time Domain Conversion with a Variant of CZT*

CZT is a conversion method that calculates the z-transform following the outline spirally or circularly from any starting point in the z-plane[16, 23]. The CZT is a flexible method to swap a group of samples to another group of points. When a variant of CZT is used to convert the collected frequency domain data to the time domain, a better distribution of the frequency samples can be obtained, and these samples are transferred to time-domain sequences [16].

A conversion operator, the variant of CZT, which transfers frequency domain data to time domain data, is defined here as

$$x_n = \frac{1}{N}\sum_{k=1}^{N} X_k \exp\left(\frac{(-\alpha + j2\pi)kn}{N}\right), \quad (1)$$

where $X_k$ is $N$ frequency-domain samples, which is the transmission coefficient $S_{21}$ recorded by the GPR system as presented in Section II. $x_n$ is the corresponding sequence of points in time domain. $\alpha$ is the frequency attenuation coefficient of the propagation medium. When $\alpha = 0$, (2) equals to the equation of the traditional inverse discrete Fourier transform (IDFT) [24], which is a special case where the attenuation is constant, and each frequency sample makes the same effect with the time. But in a realistic environment, $\alpha$ cannot be zero ($\alpha \neq 0$), and is dependent on the characteristic of the propagation medium[25].

*B. Window Determination Based on Time-Frequency Domain Analysis and Weighted Linear Regression Method*

Although (1) is used for the conversion of data from frequency to time domain, its overall computational cost is high since it takes all the frequency points for conversion. The attenuation part of (1) should vary with the characteristic of the propagation medium and frequency in order to optimize the effectiveness of (1). Therefore, the attenuation part of (1) is

$$W^{nk} = \exp\left(\frac{(-\alpha + j2\pi)kn}{N}\right). \quad (2)$$

In this paper, a frequency window, a matrix $H$ based on the magnitude of the $W^{nk}$ is proposed. For a pair of time and frequency indices $(n, k)$, given a known $\alpha$, the point $(n, k)$ in $H$ equals 0 when its corresponding magnitude of the $W^{nk}$ is less than a certain proposed value.

Therefore, a frequency to time conversion is realized as

$$x_n = \frac{1}{N}\sum_{k=1}^{K_n} X_k H_{n,k} \exp\left(\frac{j2\pi kn}{N}\right). \quad (3)$$

Here, $K_n$ is the frequency indices of the final nonzero sample of matrix $H$ at any time indices $n$. The determination of these pairs of $(K_n, n)$ is the key proposal in this work. For the same time index $n$, the element of matrix $H$ before frequency indices $K_n$ equals $\exp\left(\frac{-\alpha kn}{N}\right)$.

From the equations above, the shape of window matrix $H$ can be determined by two parameters: 1. The attenuation coefficient of the soil $\alpha$, dependent on the propagation medium; 2. The magnitude of the $W^{nk}$. Both parameters determine the shape of the window matrix $H$.

By deriving the most suitable relation between $\alpha$ and $|W^{nk}|$, the edge of the filtering window matrix in the time-frequency domain can be determined. To determine the filtering window matrix $H$, the time-frequency domain analysis is performed.

*1) Time-Frequency Domain Analysis*

After preprocessing the collected data (details in Section III), the time-frequency domain analysis was conducted to verify the frequency response of the buried targets at different depths. STFT is a widely used method to keep data information in both time and frequency domains and it has been shown to provide effective joint time-frequency analysis in two-dimensional characteristics for target detection, balance the easy implementation and good time-frequency resolution[18]. As we introduce later in Section IV, a time resolution of 2ns (20ns/10) is a proper choice for implementing our method. Other time-frequency domain analyses with high time resolution but higher complexity like empirical mode decomposition (EMD) [26], ensemble empirical mode decomposition (EEMD) [27], complete ensemble empirical mode decomposition (CEEMD) [28] based and variational-mode decomposition (VMD) based [29] are introduced, difference in their application is an interesting topic but not the focus of this paper, it can be studied in our future work.

In STFT, the local Fourier transform is implemented on time domain data that are evenly divided into small time windows, the transformed data are then stacked as a two-dimensional time-frequency plot. Mathematically, the relationship between frequency spectrum change with time is given as

$$X(\tau, \Omega) = \int_{-\infty}^{\infty} x(\tau) w(\tau - t) e^{-j\Omega\tau} d\tau, \quad (4)$$

where $x$ is the A-scan signal, $\Omega$ is the frequency. In this case, a total of 1001 ($N = 1001$) points are recorded in each A-scan trace. $\tau$ is the time step which equals 160 ps and the digitizer's sampling frequency is 6.25 Gsps. A Hamming window is applied, $w(t)$, to each data block. The window size is 1/10 of the total time indices.

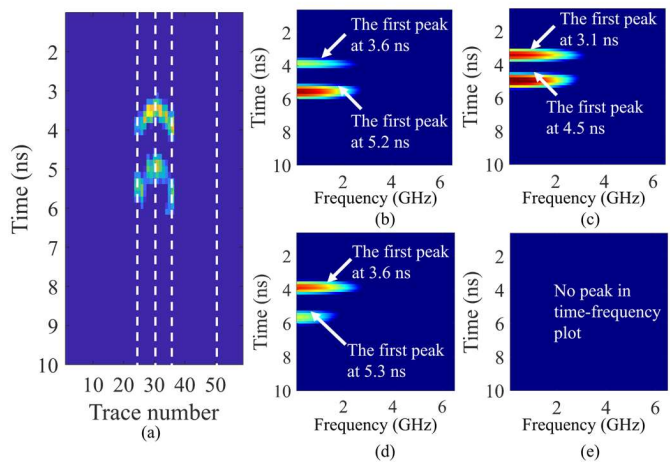

Fig. 2. (a) Processed B-scan of the real site root test, and STFT plots of traces of (b) 25th, (c) 31st, (d) 35th, (e) 50th A-scans

To illustrate the effectiveness of the STFT in the time-frequency spectrum analysis of subsurface target, a set of results from the real test data where a root is buried in the soil is given in Fig. 2. A B-scan is shown in Fig. 2(a). The STFT results of the A-scan corresponding to the chosen 25th, 31st, 35th and 50th traces are shown in Fig. 2(b)(c)(d)(e). In Fig. 2(b), two strong peaks occur at 3.6 ns and 5.2 ns on the 25th trace to the root's left side. Fig. 2(c) shows the A-scan directly above the root (the 31st trace). Therefore, the strongest peaks are found on this scan with less Electromagnetic (EM) wave transmission time at 3.1 ns and 4.5 ns. Fig. 2(d) shows strong peaks at 3.6 ns and 5.3 ns on the 35th trace to the root's right side. The time of the strongest peak of all 3 traces agrees well with the time of the received return signal from the root, the hyperbola in Fig. 2(a). The 50th trace is not in the region of interest and the STFT plot is shown in Fig. 2(e). As expected, there are no strong peaks seen. The STFT figures not only show us time domain information but also frequency domain information. From the figures, we can determine the frequency band from which there are strong reflections returned from the target. In Fig. 2(c), reflection has a wider frequency band than it has in Fig. 2(b) and 2(d). This is because at the 31st A-scan point, which is right above the root, the EM wave transmission path to the root is shorter than that at the 25th and 35th A-scan points. Note that the higher frequency samples attenuate less as well. A strong resonance in the time-frequency domain (STFT plots) indicates a reflection from the target at the corresponding time and frequency.

The STFT analysis is applied to all the A-scans data of one of the pre-processed B-scans. To better analyze the distribution of the spectrum in the time-frequency domain and make the fitted filtering solver from the B-scan testing route applicable to the whole B-scan data, all the A-scans STFT time-frequency spectrum are stacked together to form an overall STFT histogram shown in Fig. 3. There is a clear boundary in the time vs frequency domain where are points and negligible points. The curve shape of this boundary can be represented by a window $H_{n,k}$ given in (4). Fig 4 shows the occurrence counts of every frequency-time pair from the A-scans in a STFT data matrix. The occurrence of an element indicates a nonzero value in the STFT domain. The colormap in Fig. 3 indicates the number of occurrences for each time frequency pair $(n, k)$. The black region indicates no response at that time-frequency pair.

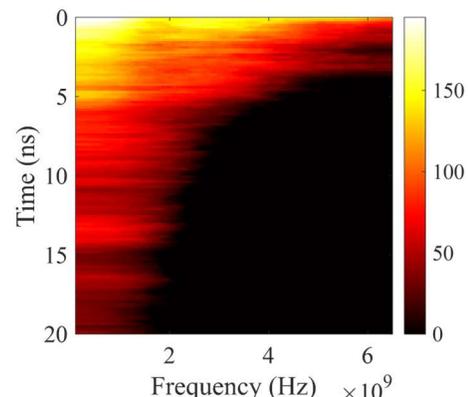

Fig. 3. Histogram of counts of points in STFT domain of the whole B-scan.

In this paper, we propose a filtering window along the boundary between the chromatic area and the black area. To obtain the boundary, a fitting technique is used to decide on the shape of the filtering window $H_{n,k}$ in (3) based on the histogram in Fig. 3. Since the time-frequency domain spectrum varies when studying different soil environments, the fitting window varies with the spectrum, which makes our method depth-adaptive to different subterrain environment.

*2) Weighted Linear Regression Method*

To determine the edge of the filtering window matrix $H_{n,k}$, a WLR method is used by adopting the points on the boundary $(n, k)$ in Fig. 3. WLR is a commonly accepted method for fitting formulas and curves [22]. The certain value that the magnitude of the $W^{nk}$ in (2) should be less than is expressed by attenuation characteristic $\alpha$ and $(n, k)$ on the boundary as

$$|W^{nk}| = \left| exp\left(-\alpha \cdot \frac{k \cdot n}{N}\right) \cdot exp\left(i \cdot 2\pi \cdot \frac{k \cdot n}{N}\right) \right|. \quad (5)$$

The relation between $\alpha$ and $|W^{nk}|$ can be obtained from (5), where the pairs of $(n,k)$ lie along the boundary that satisfies

$$\frac{\alpha}{ln|W^{nk}|} + \frac{N}{k \cdot n} = 0. \quad (6)$$

Let $\gamma = \alpha/ln|W^{nk}|$, for $ln|W^{nk}| \neq 0$. Then a suitable constant, $\gamma$, can be found such that the equation,

$$\gamma + \frac{N}{k \cdot n} = 0, \quad (7)$$

is satisfied. According to the number of elements occurring in the different time intervals, the element at different times has varying contributions to determining the trend of the shape of the window. In both low and high-frequency samples, there are much more elements in the early time samples than in the late time samples. Therefore, different pairs of data should be given different weights.

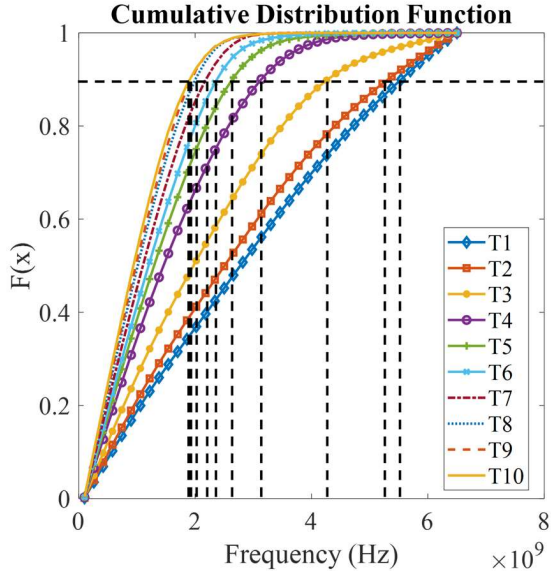

Fig. 4. CDFs of the number of elements at 10 equal time intervals (T) in the B-scan STFT matrix, each time interval is 2 ns. CI = 0.9 is labelled in dot line.

The cumulative distribution functions (CDFs) of the number of elements at the different time intervals in the B-scan STFT matrix are drawn in Fig. 4. The confidence interval (CI) of 0.9 and 0.95 are commonly used in statistics [30]. Through a series of iterative experiments, we found 0.9 to be the optimal choice for our scenario in order to fulfill the selection of the frequency sample indices $k_i$ corresponding to each time interval index $n_i$ in Fig. 4.

We aim to have a curve that fits the boundary that satisfies (7) and has adopted the WLR method to do so. To determine the boundary curve, a suitable $\gamma$ should be determined via

$$\min_{(\gamma)} \sum_{i=1}^{m} \left(\gamma + \frac{N}{k_i n_i}\right)^2 \cdot \omega_i. \quad (8)$$

In (8), $n_i$ is the time indices, and $k_i$ is the frequency indices. $\omega_i$ is the weight, which is determined by the number of elements in the B-scan STFT matrix at a time interval $n_i$. $m$ is the total number of pairs of $(n_i, k_i)$, in this illustration, $m$ is chosen as ten equal time intervals, the selection of $m$ is detailed with the data analysis given in Section IV.

Since it is a quadratic optimization problem, it has a unique solution given as

$$\hat{\gamma} = -\frac{\sum_{i=1}^{m} \frac{N}{k_i n_i} \cdot \omega_i}{\sum_{i=1}^{m} \omega_i}. \quad (9)$$

The relation between $\alpha$ and $|W^{nk}|$ can be determined and the value of $\gamma$ is derived. The result is then used to fit the boundary curve of the time-frequency filter window $H_{n,k}$. The fitted curves differ depending on the test site soil scenarios. A comparison of the filtering windows for different soil scenarios is given in Section IV.B.

The frequency window $H_{n,k}$ can be shaped based on the window curve fitted based on (9). After applying the collected frequency-domain data into (3), the data is transferred to time domain with the adaptive signal filtering. How the signal filtering is adaptive is demonstrated by comparing two GPR survey examples in different scenarios in Section IV A.

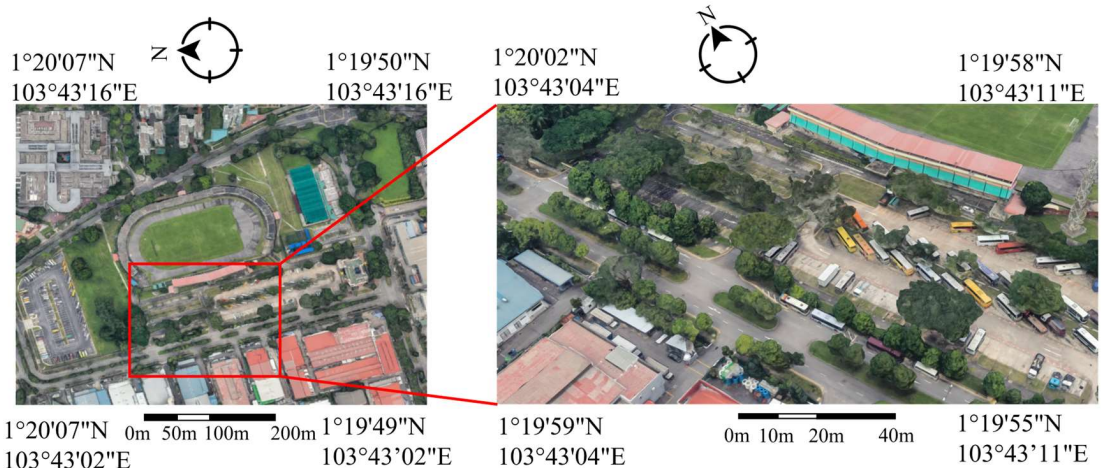

Fig. 5. Test site for the GPR tree root detection (map data: Google).

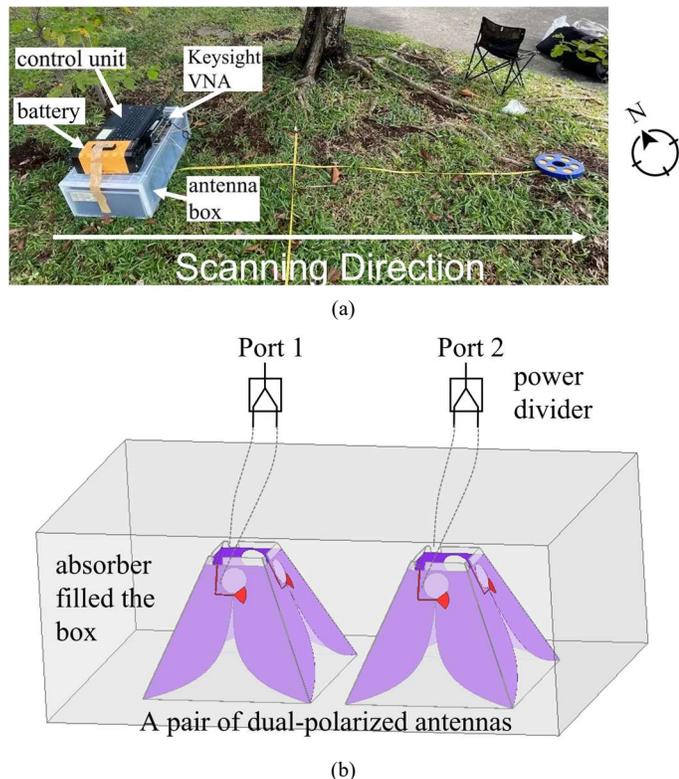

Fig. 6. (a) Schematic of the experiment scenario and the view of the real scenario. (b) The two dual-polarized antennas presented in [31] work as the transceiver.

## III. DATA ACQUISITION AND DATA PRE-PROCESSING

The tree root detection study was conducted near Chin Bee Road, a typical roadside tree environment, in Singapore (Fig. 5) on the 3rd of November 2021. An Angsana tree was under investigation and identified for this study by the National Parks Boards (N-parks), Singapore.

The GPR surveys were carried out along straight lines on one side of the tree trunk. A set of 5 straight line scans were performed. The distance between each straight-line scan is 0.5 m. The first B-scan test was conducted 1.0 m from the bark of the tree; it allows enough space for the GPR testing and avoids the interference of visible tree roots exposed above ground. 200 A-scans were recorded by moving the GPR system along the preset scanning trace with a step size of 0.03 m. The data was further processed using preprocessing techniques. Fig. 6 shows the GPR survey setup.

The SFCW UWB GPR system consists of a computer-controlled Keysight P5008A Vector Network Analyzer (VNA) as a transceiver and two compact dual-polarized Vivaldi antennas [31], as shown in Fig. 6. Two antennas are sealed in a box and are separated 0.1 m apart. Absorbers are placed around the antennas to reduce the direct coupling between the antennas and environmental noise. The box is moved along the ground with the height of antennas above the soil surface of about 0.01 m. The VNA sweeps 1001 frequency points in the frequency band from 0.25 GHz to 6.5 GHz; this band is usually chosen for subsurface scanning [17], and the corresponding frequency step size is 6.25 MHz. When the relative soil permittivity ($\varepsilon_r$) is about 10, the corresponding depth resolution of the set frequency band can be estimated using

$$\Delta\delta = \frac{1}{2 \times BW} \cdot \frac{c_0}{\sqrt{\varepsilon_r}}, \quad (10)$$

where $c_0$ is the speed of the EM wave in air and $BW$ is the system bandwidth ($BW = 6.25$ GHz). The calculated depth resolution is about 0.01 m. The depth resolution is small relative to most of the roots' diameters, which means that the system settings are suitable for the detection of root systems, including thin roots. The time window of the system setting is about 80 ns, and the corresponding penetration depth is about 10 m (this depth varies depending on the soil properties). Taking into consideration the tropical soil properties and the depth of the tree roots of interest to be less than 2.5 m, the former 20 ns time window is chosen in this study. The intermediate frequency bandwidth (IFBW) is set as 1 kHz, and the power is set as -10 dBm.

In the measurement, the two opposing Vivaldi elements of the two dual-polarized antennas, which are almost parallel to the growth direction of the tree roots (Fig. 5), are combined using a power divider, and then connected to port 1 and port 2 of the VNA, respectively. The transmission coefficient $S_{21}$ is recorded for processing. Before executing experiments, a calibration procedure is necessary, and in this work, we employ a Keysight electronic calibration (E-cal) module: 85093C. A standard 'full two-port calibration' is performed.

The data collected goes through preprocessing in order to obtain a relatively clean B-scan image [2, 4]. The basic preprocessing includes

1. DC removal to get a GPR trace with an average amplitude different from zero, prevent GPR signals from being distorted by initial direct current (DC) shifts.

2. Time zero correction to fix the starting time reference of the GPR data, make sure the reflection time can accurately indicate the depth of the target.

3. Background removal using singular value decomposition (SVD) to separate ringing noise from the response from the targets, proved effective in discrimination between two even slightly different images [17].

## IV. MEASUREMENT DATA ANALYSIS

### A. Time-Frequency Window Outcome and Soil-information-independent Characteristic of the DATFF

The effect of the number of time-frequency samples $m$ in (8) and (9) on the shape of the fitted curve, value of $\gamma$, when applying the proposed method to the Angsana tree GPR B-scan data, is shown in Fig. 7. As shown, the value of $m = 10$ and above, the value of $\gamma$ levels out. To balance between stability and computation cost, $m = 10$ is used in our experiment.

The fitted window curve of the B-scan STFT matrix (Fig. 3) collected from the Angsana tree is shown in Fig. 8(a), the $\gamma$ value is -0.058 in this case. The histograms of counts of points, and the fitted window curve between two different B-scan data of the same Angsana tree detection (Fig. 6) are shown in Fig. 8 (a)(b). From the figures, we can see the characteristics of two B-scan data from the same tree detection have a similar trend and the fitted window curves are the same. This implies that the derived filtering window for the soil region around the same tree will be the same.

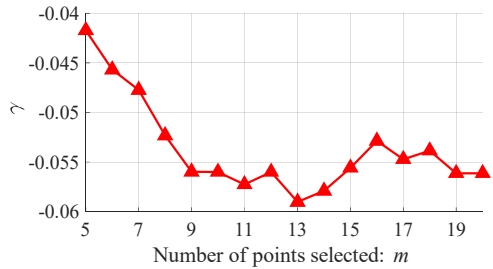

Fig. 7. Illustration of effect of m in (9) on the values of γ when applying the proposed method on a GPR B-scan data.

The proposed data processing method is also applied on measurement results collected from a tree at Nanyang Technological University (NTU) (1°20'28"N, 103°40'51"E) on 26th of July. The GPR system setting is the same as the Angsana tree measurement, and a total of 101 A-scans of each B-scan were collected. Similar to the Angsana tree, a total of 5 B-scans were collected.

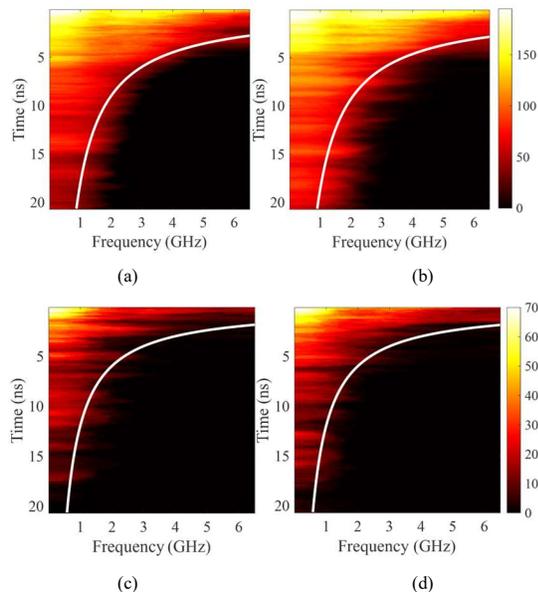

Fig. 8. Fitted window curve of (a)(b): two different B-scan data of the same Angsana tree detection, (c)(d): two different B-scan data of the same tree detection in NTU.

After joint time-frequency domain analysis of the collected data, Fig. 8(c)(d) shows two histograms of counts of points in STFT matrices for 2 B-scans from the NTU tree root detection. In Fig. 8(c)(d), the two fitted window curves have the same shape. However, the filter window derived from the tree at NTU is different from that derived from the Angsana tree. This shows that the filter window cannot be generalized and is dependent on the soil environment surrounding the tree. Comparing the soil environment surrounding the tree at NTU and the Angsana tree, the frequency band for frequency-time conversion in any time interval is narrower for the tree at NTU, which means more frequency responses at higher-frequency bands are filtered out. The different shape of the window curve means different relation between $\alpha$ and $|W^{nk}|$, In Fig. 9, we compare the values of $\gamma$ of 5 B-scans data from both tree surveys at the two different testing sites. We can see the $\gamma$ of the Angsana tree is greater than the $\gamma$ of the tree in NTU. The variation of the value of $\gamma$ is slight and can be ignored in the same tree soil environment. The two surveys were performed on different days with different weather conditions. The Angsana tree was surveyed during a relatively dry day during noon time. Therefore, the relative humidity on that day is comparably low, at about 65% to 70%. The tree at NTU was surveyed during a relatively wet day when the humidity was over 80% [32].

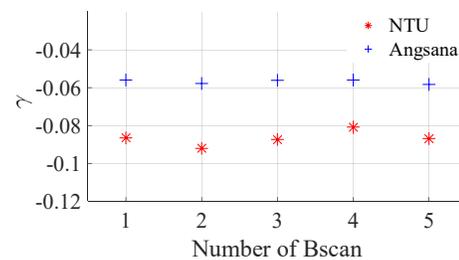

Fig. 9. Comparison of the values of $\gamma$ of 5 B-scans data from two trees surveys in two different sites.

B. *DATFF Performance Evaluation*

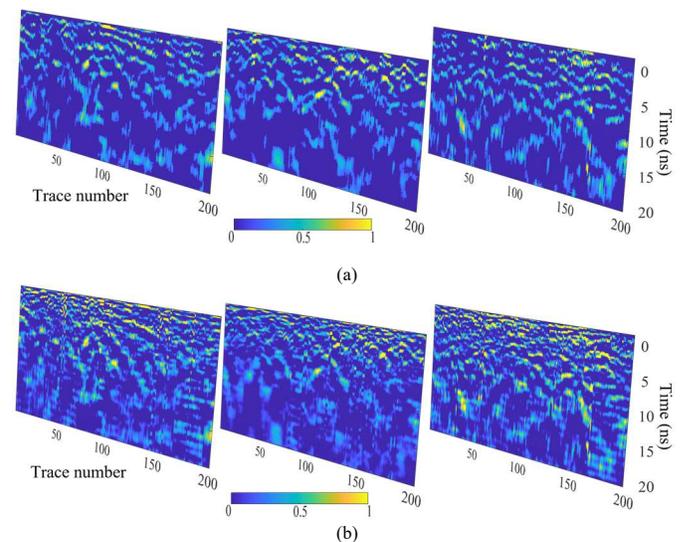

Fig. 10. Three B-scans images processed by (a) the proposed DATFF methodology and (b) ISDFT.

Images of three normalized B-scans conducted for an Angsana tree, then processed through DATFF and ISDFT method with a pre-set constant $\alpha = 0.01$, equivalent $|W^{nk}|$ equals to 0.5 [16] are shown in Fig. 10. All three images in Fig. 10(a) have better quality than their counterpart in Fig. 10(b). Hyperbolae in Fig. 10(a) can be better identified; less noise of non-interest is in Fig. 10(a) compared with Fig. 10(b). Meanwhile, the resolution of the images in Fig. 10(a) is better than that in Fig. 10(b). From B-scans in Fig. 10(a), two adjacent hyperbolae that represent neighbouring roots can be visually identified, while in Fig. 10(b), most of these adjacent hyperbolic signatures are merged together as one signature. These real data scans illustrate the potential of DATFF for higher interpretation accuracy and lower noise, especially within the time range of [0 ns, 10 ns], the corresponding depth range is between [0 m, 1.25 m], where tree roots are commonly found.

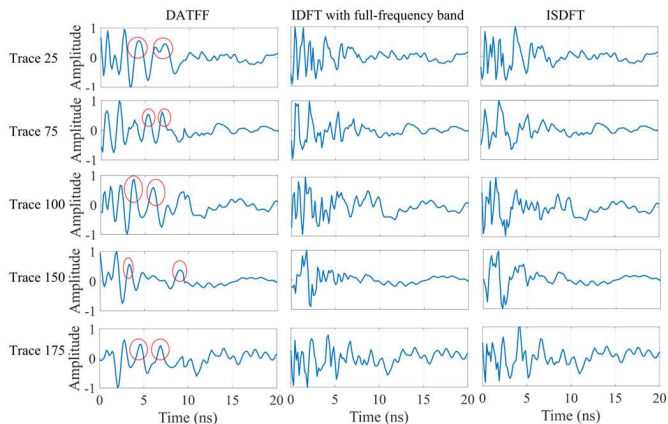

Fig. 11. A-scans comparison between three frequency-to-time-domain method of the A-scan trace no. 25, 75, 100, 150, 175 in the third B-scan of Fig. 10.

To show the efficiency of the DATFF method in detail, the third B-scan is used as an example. Fig. 11 shows the time domain A-scans corresponding to different scanning points of this B-scan trace taken from the three different post-processing. It is observed that the proposed DATFF effectively suppresses background noise, emphasizes the peaks of the root targets (as circled out), smooths out the noise of the A-scan curve, and therefore, improving the signal-to-clutter ratio (SCR).

The third time-domain B-scan image of the Angsana tree processed through our proposed methodology is detailed in Fig. 12(c). For comparison, the time domain B-scan image processed through ISDFT method is shown in Fig. 12(d). From the two figures, we can see that the hyperbolae indicating different roots are clearly shown in Fig. 12(c). Below 5 ns, the proposed filtering window has removed a significant amount of noise. Above 5 ns, although less noise is removed, the hyperbolae are easily distinguishable as a result of the proposed filtering window.

In Fig. 12(c), some intersecting hyperbolae representing two roots close to or crossing each other are shown in the regions of interest (ROIs) ROI 4 and 7. In the noisy image in Fig. 12(d), the two intersecting hyperbolae cannot be distinguished and are merged together. The noise level of Fig. 12(c) is much less than Fig. 12(d) as a result of the proposed filtering window that filters out the frequencies that contain mainly noise components. The reflected signals (noise components) at these frequency bands are at an energy level that is less than the noise level of the receiver.

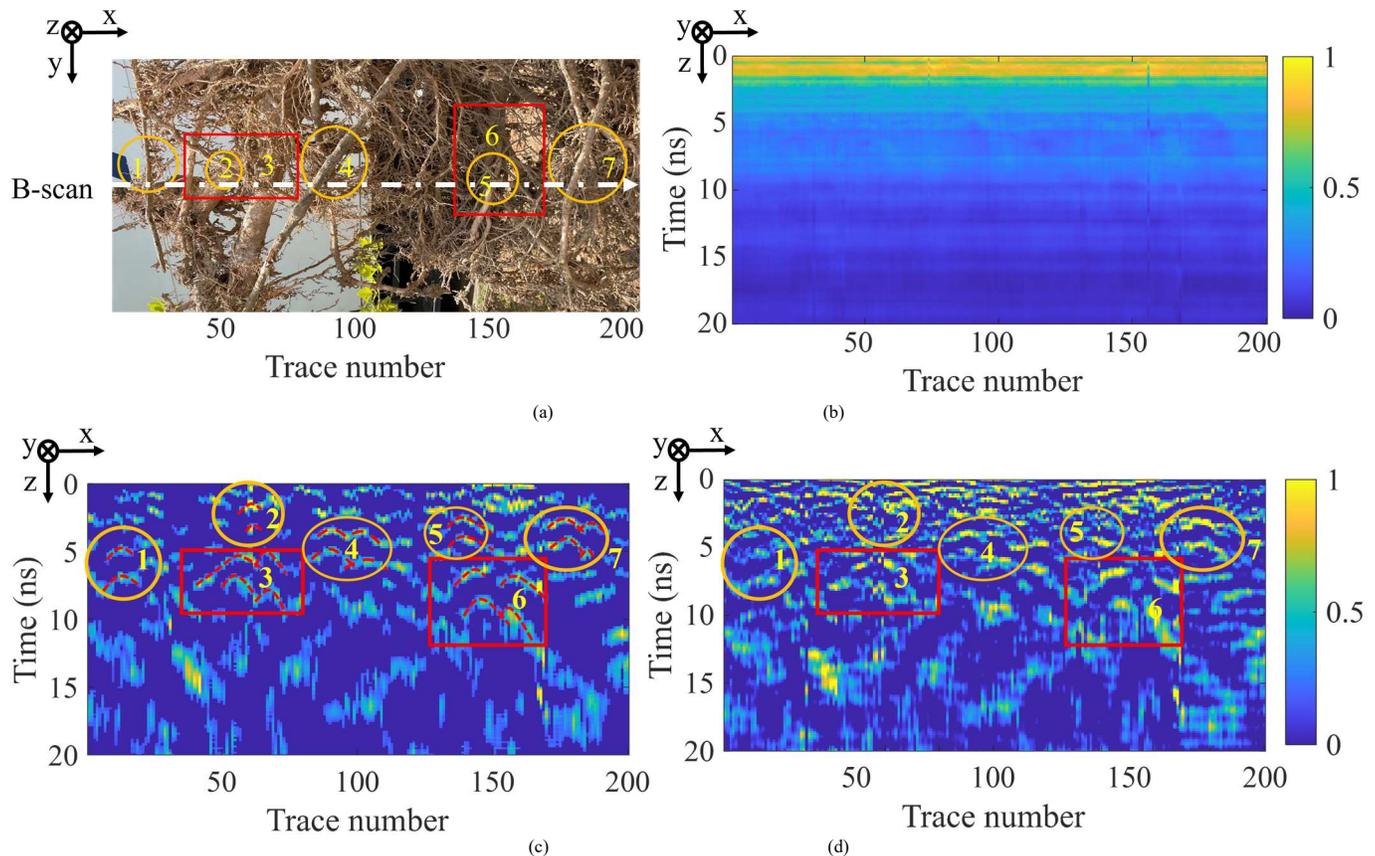

Fig. 12. Comparison between the (a) real roots under detection, (b) Raw data B-scan, (c) B-scan after the proposed DATFF methodology and (d) B-scan after the ISDFT.

To verify the effectiveness of our proposed method, the investigated tree's roots were dug out after the measurement. A comparison between the processed B-scan image and real roots configuration under the detection line is made in Fig. 12. The B-scan is conducted along the white line in Fig. 12(a), which is one of the B-scans conducted in Section II, the distance between the tree's bark to this B-scan is 2 m. The raw data image of this B-scan is drawn in Fig. 12(b), it is difficult to distinguish the hyperbolae from the roots targets due to the clutters from ringing noises of the air-to-ground interface and the mutual coupling of two antennas. This is especially so for the shallow tree roots. In Fig. 12(a) and (c), the roots and hyperbolae radargrams agree well with each other. Specifically, shallower thin roots are circled in yellow and two deeper thick main roots are framed in red, the two main roots, ROI (root) 6 is deeper than ROI (root) 3 and below ROI (root) 5. ROI 4 and 7 in Fig. 12(a), both consist of 2 adjacent hyperbolae, which correspond to two cross roots, as shown in Fig. 12(c).

In addition, to quantify the performance of the DATFF method, the signal to clutter ratio (SCRs) for the different processing methods is compared. SCR is calculated as

$$SCR = 10\,log\left(\frac{N_c}{N_I}\frac{\sum_{p\in R_I}|v(p)|^2}{\sum_{p\in R_c}|v(p)|^2}\right), \quad (11)$$

where $v(p)$ is the value of the $p$-th pixel in the B-scan image. $N_c$ and $N_I$ are the number of pixels in the clutter region $R_c$ and region of interest $R_I$, respectively. The higher the SCR, the better the processing technique (in the case of the proposed DATFF technique, the better the filtering window). In this paper, signals are the ROI corresponding to roots' ground truth in the B-scan, and clutters are the regions of non-interest with no roots' ground truth in the B-scan.

Table II lists the quantitative analysis of different processing methods in terms of SCR for the B-scan image of the Angsana tree. On the base of full-frequency band + SVD (usually used in SFCW GPR processing) with a SCR of -0.3 dB, the proposed method achieves the best SCR of 0.10 dB. The ISDFT proposed in [16] with a pre-set constant $\alpha = 0.01$, equivalent $|W^{nk}|$ equals to 0.5, gives an SCR of -0.06 dB. This shows the importance of using our proposed 'soil-information-independent' depth-adaptive filtering window. The method in [15] is better than the full-frequency band + SVD, but worse than our DATFF method. The SCR results of the full-frequency band and DATFF with another background removal method 'mean subtraction (MS)' are also listed. It is clear that DATFF also gives a better result of -1.90 dB than the full-frequency band result of -3.36 dB. Meanwhile, the background removal method of SVD works better than MS in this paper, this is consistent with the results presented in [33].

TABLE II
SCRs FOR PROCESSING METHODS ON ANGSANA B-SCAN DATA

| Processing methods | SCRs (dB) |
|---|---|
| IDFT with Full-frequency band + SVD | -0.3 |
| ISDFT + SVD | -0.06 |
| DATFF + SVD | 0.10 |
| IDFT with Full-frequency band + MS | -3.36 |
| ISDFT + MS | -2.82 |
| DATFF + MS | -1.90 |

## V. CONCLUSION

This paper has demonstrated a depth-adaptive frequency-selective processing method called DATFF, which is an integration of frequency-time filtering and a variant of CZT, to enhance SFCW GPR tree roots B-scans without the knowledge of the transmission medium. The frequency-time filtering method can be effectively obtained from a joint time-frequency analysis method STFT and WLR technology. The proposed method is successfully applied to a group of real test data collected from tropical areas. The B-scan processed by the proposed method clearly shows the hyperbolae with low noise. To validate the processed result, the roots of the tree under test are dug out; tree roots under the corresponding B-scan trace agree well with the two-dimensional (2D) hyperbolic features in the processed B-scan. By applying the method to the data collected in different climate situations and test sites, we can conclude that the DATFF is a soil-information-independent tool that is not limited by the unknown characteristic of the propagation medium. From the processed result and its comparison with ground-truth, the proposed filtering method is demonstrated to be a promising technique for signal-to-noise ratio improvement under tropical area tree root detection using SFCW GPR. Subsequent B-scan images with lower noise levels can facilitate further research on GPR, such as hyperbolic feature segmentation, health monitoring of each root, and classification of the subsurface targets.

## REFERENCES

[1] A. Aboudourib, M. Serhir, and D. Lesselier, "A processing framework for tree-root reconstruction using ground-penetrating radar under heterogeneous soil conditions," *IEEE Transactions on Geoscience and Remote Sensing,* vol. 59, no. 1, pp. 208-219, 2020.

[2] W. Luo, Y. H. Lee, H.-H. Sun, L. F. Ow, M. L. M. Yusof, and A. C. Yucel, "Tree Roots Reconstruction Framework for Accurate Positioning in Heterogeneous Soil," *IEEE Journal of Selected Topics in Applied Earth Observations and Remote Sensing,* vol. 15, pp. 1912-1925, 2022.

[3] H.-H. Sun, Y. H. Lee, W. Luo, L. F. Ow, M. L. M. Yusof, and A. C. Yucel, "Dual-cross-polarized GPR measurement method for detection and orientation estimation of shallowly buried elongated object," *IEEE Transactions on Instrumentation and Measurement,* vol. 70, pp. 1-12, 2021.

[4] L. Lantini, F. Tosti, I. Giannakis, L. Zou, A. Benedetto, and A. M. Alani, "An enhanced data processing framework for mapping tree root systems using ground penetrating radar," *Remote Sensing,* vol. 12, no. 20, p. 3417, 2020.

[5] W. Luo, Y. H. Lee, L. F. Ow, M. L. M. Yusof, and A. C. Yucel, "Accurate Tree Roots Positioning and Sizing Over Undulated Ground Surfaces by Common Offset GPR Measurements," *IEEE Transactions on Instrumentation and Measurement,* vol. 71, pp. 1-11, 2022.

[6] O. Brandt, K. Langley, J. Kohler, and S.-E. Hamran, "Detection of buried ice and sediment layers in permafrost using multi-frequency Ground Penetrating Radar: A case examination on Svalbard," *Remote Sensing of Environment,* vol. 111, no. 2-3, pp. 212-227, 2007.

[7] F. Soldovieri and L. Orlando, "Novel tomographic based approach and processing strategies for GPR measurements using multifrequency antennas," *Journal of Cultural Heritage,* vol. 10, pp. e83-e92, 2009.

[8] L. Liu and L. Zhu, "GPR signal analysis: can we get deep-penetration and high-resolution simultaneously?," in *Proceedings of the Tenth International Conference on Grounds Penetrating Radar, 2004. GPR 2004.*, 2004, vol. 1: IEEE, pp. 263-265.

[9] J. Xiao and L. Liu, "Permafrost subgrade condition assessment using extrapolation by deterministic deconvolution on multifrequency GPR data acquired along the Qinghai-Tibet railway," *IEEE Journal of Selected Topics in Applied Earth Observations and Remote Sensing,* vol. 9, no. 1, pp. 83-90, 2015.

[10] W. Zhao and G. Lu, "A novel multi-frequency GPR data fusion algorithm based on time-varying weighting strategy," *IEEE Geoscience and Remote Sensing Letters,* 2021.
[11] J. Leckebusch, "Comparison of a stepped‐frequency continuous wave and a pulsed GPR system," *Archaeological Prospection,* vol. 18, no. 1, pp. 15-25, 2011.
[12] D. Seyfried and J. Schoebel, "Stepped-frequency radar signal processing," *Journal of applied geophysics,* vol. 112, pp. 42-51, 2015.
[13] E. S. Eide and J. F. Hjelmstad, "3D utility mapping using electronically scanned antenna array," in *Ninth International Conference on Ground Penetrating Radar*, 2002, vol. 4758: International Society for Optics and Photonics, pp. 192-196.
[14] F. Lombardi and M. Lualdi, "Step-frequency ground penetrating radar for agricultural soil morphology characterisation," *Remote Sensing,* vol. 11, no. 9, p. 1075, 2019.
[15] J. Sala and N. Linford, "Processing stepped frequency continuous wave GPR systems to obtain maximum value from archaeological data sets," *Near Surface Geophysics,* vol. 10, no. 1, pp. 3-10, 2012.
[16] J. Sala, H. Penne, and E. Eide, "Time-frequency dependent filtering of step-frequency ground penetrating radar data," in *2012 14th International Conference on Ground Penetrating Radar (GPR)*, 2012: IEEE, pp. 430-435.
[17] T. G. Savelyev, L. Van Kempen, H. Sahli, J. Sachs, and M. Sato, "Investigation of time–frequency features for GPR landmine discrimination," *IEEE Transactions on Geoscience and Remote Sensing,* vol. 45, no. 1, pp. 118-129, 2006.
[18] I. L. Al-Qadi, W. Xie, D. L. Jones, and R. Roberts, "Development of a time–frequency approach to quantify railroad ballast fouling condition using ultra-wide band ground-penetrating radar data," *International Journal of Pavement Engineering,* vol. 11, no. 4, pp. 269-279, 2010.
[19] Y. Zhang, P. Candra, G. Wang, and T. Xia, "2-D entropy and short-time Fourier transform to leverage GPR data analysis efficiency," *IEEE Transactions on Instrumentation and Measurement,* vol. 64, no. 1, pp. 103-111, 2014.
[20] L. Lantini, F. Tosti, L. B. Ciampoli, and A. M. Alani, "A frequency spectrum-based processing framework for the assessment of tree root systems," in *SPIE Future Sensing Technologies*, 2020, vol. 11525: International Society for Optics and Photonics, p. 115251J.
[21] L. Lantini, F. Tosti, L. Zou, L. B. Ciampoli, and A. M. Alani, "Advances in the use of the Short-Time Fourier Transform for assessing urban trees' root systems," in *Earth Resources and Environmental Remote Sensing/GIS Applications XII*, 2021, vol. 11863: SPIE, pp. 212-219.
[22] S. Weisberg, *Applied linear regression*. John Wiley & Sons, 2005.
[23] L. R. Rabiner, R. W. Schafer, and C. M. Rader, "The chirp z‐transform algorithm and its application," *Bell System Technical Journal,* vol. 48, no. 5, pp. 1249-1292, 1969.
[24] D. Sundararajan, *The discrete Fourier transform: theory, algorithms and applications*. World Scientific, 2001.
[25] M. Bano, "Modelling of GPR waves for lossy media obeying a complex power law of frequency for dielectric permittivity," *Geophysical Prospecting,* vol. 52, no. 1, pp. 11-26, 2004.
[26] A.-O. Boudraa and J.-C. Cexus, "EMD-based signal filtering," *IEEE transactions on instrumentation and measurement,* vol. 56, no. 6, pp. 2196-2202, 2007.
[27] T. Wang, M. Zhang, Q. Yu, and H. Zhang, "Comparing the applications of EMD and EEMD on time–frequency analysis of seismic signal," *Journal of Applied Geophysics,* vol. 83, pp. 29-34, 2012.
[28] J. Li, C. Liu, Z. Zeng, and L. Chen, "GPR signal denoising and target extraction with the CEEMD method," *IEEE Geoscience and Remote Sensing Letters,* vol. 12, no. 8, pp. 1615-1619, 2015.
[29] Y.-J. Xue, J.-X. Cao, D.-X. Wang, H.-K. Du, and Y. Yao, "Application of the variational-mode decomposition for seismic time–frequency analysis," *IEEE Journal of Selected Topics in Applied Earth Observations and Remote Sensing,* vol. 9, no. 8, pp. 3821-3831, 2016.
[30] F. D. C. Kraaikamp and H. L. L. Meester, "A Modern Introduction to Probability and Statistics," ed: Springer: Berlin/Heidelberg, Germany, 2005.
[31] H.-H. Sun, Y. H. Lee, W. Luo, L. F. Ow, M. L. M. Yusof, and A. C. Yucel, "Compact Dual-Polarized Vivaldi Antenna with High Gain and High Polarization Purity for GPR Applications," *Sensors,* vol. 21, no. 2, p. 503, 2021.
[32] M. S. Singapore. "Climate of Singapore." http://www.weather.gov.sg/climate-climate-of-singapore/ (accessed December 28, 2021.
[33] A. De Coster and S. Lambot, "Full-wave removal of internal antenna effects and antenna–medium interactions for improved ground-penetrating radar imaging," *IEEE Transactions on Geoscience and Remote Sensing,* vol. 57, no. 1, pp. 93-103, 2018.